\documentclass[aps,prb,twocolumn,showpacs,superscriptaddress]{revtex4}
\usepackage{graphicx}
\usepackage{bm}
\newcommand{\beq}{\begin{equation}}
\newcommand{\eeq}{\end{equation}}

\begin{document}

\title{The low-temperature collapse  of the Fermi surface
 and phase transitions  in  correlated  Fermi systems}
\author{V.~A.~Khodel}
\affiliation{Russian Research Centre Kurchatov
Institute, Moscow, 123182, Russia}
\affiliation{McDonnell Center for the Space Sciences \&
Department of Physics, Washington University,
St.~Louis, MO 63130, USA}
\date{\today}
\begin{abstract}
A topological crossover, associated with the collapse of the Fermi surface  in strongly
correlated Fermi systems, is examined. It is  demonstrated that in these systems, the  temperature
 domain where standard Fermi liquid results hold dramatically narrows, because  the Landau regime
  is replaced by a classical one. The impact of the  collapse of the
Fermi surface on pairing correlations is analyzed. In the domain of the Lifshitz phase diagram
where the Fermi surface collapses, splitting of the BCS superconducting  phase transition into
two different ones of  the same symmetry is shown to occur.

\end{abstract}

\maketitle

{\it Introduction}.
For  the past few years, the investigation of topological transitions
in  correlated Fermi systems that  dates back to a pioneer work  by I. M. Lifshitz,
published in 1960,\cite{lifshitz} has become one of  hot topics in condensed
 matter physics.\cite{ndl,tai1,prb2008,norman,vol2011b,vol2011c,vol2011d,kopnin,tai2,shag2011,pank2011,jetplett2011,schnyder,brudon}
  These transitions are responsible for non-Fermi-liquid (NFL) behavior that
 manifests itself in  singularities of thermodynamic characteristics of Fermi
 systems. E.g. in  FL theory,  the spin susceptibility  $\chi(T)$ remains unchanged
 at $T\to 0$. Contrariwise, at the transition point, $\chi(T\to 0)$ diverges as $T^{-\alpha}$,
  with the critical index $\alpha>0$, somewhat depending   on the shape of the
  single-particle spectrum $\epsilon({\bf p})$.\cite{prb2005,yak,shag2011}
In case the function   $\epsilon({\bf p})$ is measured from the chemical potential $\mu$,
 the topological  rearrangement of the Landau  state is associated with  the change in
   the number     of its zeroes.\cite{lifshitz,volrev}
   In conventional  nonsuperfluid homogeneous  Fermi systems, whose Fermi surfaces are
       singly connected,  equation
  \beq
 \epsilon(p)=0
 \label{topeq}
 \eeq
  has   the single {\it real} root, the   Fermi momentum $p_F$. In the original article,
  Lifshitz analyzed topological transitions, occurring in noninteracting electron gas  of
   metals at high pressures.    Scenarios for such transitions, entailed by interactions
   between quasiparticles,   one of which is  addressed in this article, have emerged
 thirty years later,\cite{ks,vol,noz,zb,shagp}  (for a recent  review, see   Ref.\onlinecite{shagrep}).
 NFL behavior of Fermi systems near  one of the critical points of Eq.(\ref{topeq}) associated
  with the divergence of the effective mass $M^*$, (this point is called the quantum critical point (QCP)),
    has been studied extensively in recent years.\cite{loh,stegl}

 Usually Eq.(\ref{topeq}) is analyzed at zero temperature. However, the spectrum
 $\epsilon(p)$  depends  on $T$, being a functional of the quasiparticle momentum
  distribution  $n(p,T)$ that has the standard Fermi-like form
  \beq
 n(p,T)=\left( 1+e^{{\epsilon(p)\over T}}\right)^{-1}  \ ,
 \label{dist}
 \eeq
  normalized in 3D by ordinary condition
  \beq
  \int n(p)d\upsilon=\rho \equiv {p^3_F\over 3\pi^2} \  ,
  \label{norm}
  \eeq
   with the volume   element in 3D momentum space  $d\upsilon=p^2dp/\pi^2$.

The key point of this article is that at $T>0$ where topological transitions are
  replaced by     crossovers, there is a different route of the topological
rearrangement, associated with the collapse of the Fermi surface. The collapse  occurs
 in case Eq.(\ref{topeq})  has no  real roots at all. A classical  Maxwell  reconstruction
 of the $T=0$ ideal-Fermi-gas
momentum distribution is the  simplest example of such a topological rearrangement.
In this case where $\epsilon(p)=p^2/2M-\mu(T)$,  the  trajectory  of the single root of
 Eq.(\ref{topeq}), denoted further $p_{1/2}$ due to relation  $n(p_{1/2})=1/2$, stemming
from Eq.(\ref{dist}) at $\epsilon(p)=0$,  is easily  traced. At $T\to 0$, one has
$p_{1/2}(0)=p_F$. With the temperature
  rise,  $ p_{1/2}(T)$ moves toward the origin, attaining it at temperature
 where $\mu(T)$ changes sign. At higher $T$,  real roots of  Eq.(\ref{topeq}) no
  longer exist, and the function $\epsilon(p)$ becomes positive defined,
  the  gap  in the spectrum $\epsilon(p)$ increasing with the further increase
  of $T$ that  entails rapid spread  of the  quasiparticle momentum  distribution $n(p,T)$.

  In systems with weak and/or  moderate correlations, such an option is of no
  interest,    since  temperature $T_M$, at which the Maxwell-like crossover takes place,
    is  comparable with the   Fermi   energy    $\epsilon^0_F=p^2_F/2M$.  However,
     with  strengthening correlations,  the effective mass $M^*$ increases,
and correspondingly, the FL     spectrum    $\epsilon(p)=p_F(p-p_F)/M^*$ becomes flatter
 and  flatter.  As a result,  temperature  $T_M$ goes down that leads, in its turn, to the
  dramatic shrinkage of the temperature  region  where the standard  FL  behaviors:
 $C(T)\propto T$, for the specific heat,  and $\chi(T)=const$, for the spin susceptibility,
  hold.  Importantly, at $T\simeq T_M\ll \epsilon^0_F$, the Landau-Migdal quasiparticle
  pattern remains applicable  to evaluation of {\it thermodynamic} properties.
 This   allows one  to apply the quasiparticle formalism to the investigation of
  relevant problems,  like splitting of the BCS  superconducting phase transition into
   two ones, (see below), notwithstanding   discrepancies  between experimental data
    and standard  FL predictions.

{\it Simple  model of the collapse of the Fermi surface at low $T$.} Within the
 quasiparticle picture, all the  thermodynamic properties of correlated
  Fermi  systems are  evaluated in terms of the single-particle spectrum  $\epsilon(p)$.
 This spectrum can be calculated   on the   base  of FL equation, connecting
 $\epsilon(p)$  with  the  FL  quasiparticle    momentum   distribution (\ref{dist})
 in terms of the  first harmonic $f_1$ of the  interaction function $f$,
 taken as a phenomenological input.  In the 3D case, this equation has the form:
 \beq
 {\partial\epsilon(p)\over \partial  p}={ p\over M}+{1\over 3}\int f_1( p, p_1){\partial n(p_1)\over \partial  p_1}d\upsilon_1 \  .
 \label{lansp}
 \eeq

The interaction function $f({\bf p},{\bf p}_1)$ is known to coincide with
 a specific limit of the  scattering amplitude
$\Gamma({\bf p}_1,{\bf p}_2;{\bf q}, \omega)$ where ${\bf q}={\bf p}-{\bf p}_1\to 0,\omega\to 0; q/\omega\to 0$.
 Generally $\Gamma$ contains two  different components. The first that  prevails in the vicinity of second--order
   phase transitions changes rapidly near the Fermi surface.
 The second      varies  smoothly until momenta  reach values
 much larger than $p_F$.  Neglecting
 irregular components,  the scattering amplitude $\Gamma$ can be
 written   in the standard  form $\Gamma\propto 1/(a^{-1}-r_eq^2/2)$
  where $a$ is the  scattering length, and $r_e$, the
  effective range.  This  form has to be  supplemented by   an exchange term to yield
  \beq
\Gamma({\bf p},{\bf p}_1,{\bf q})={4\pi\over M}\left({a\over 1{-}ar_eq^2/2}{-}{a/2\over 1{-}ar_e({\bf p}{-}{\bf p}_1{+}{\bf q})^2/2}\right)  \  .
\label{gam}
\eeq
  To facilitate the analysis we expand this expression into the Taylor  series.
Retaining    only two first terms in this expansion, inherent in an
effective mass  approximation, where
\beq
f({\bf p},{\bf p}_1)=\Gamma({\bf p},{\bf p}_1,{\bf q}=0)=
{2\pi a\over M}\left(1 -{ar_e\over 2}({\bf p}-{\bf p}_1)^2\right) \  ,
\label{te}
\eeq
the group velocity     $d\epsilon(p)/ d p$ is
evaluated from Eq.(\ref{lansp}) in the closed form:
  \beq
  {d\epsilon(p)\over d p}={p\over M}\left(1-2\pi a^2r_e\rho\right) \  .
  \label{ameff}
  \eeq
  The spectrum  $\epsilon(p)$ is then calculated straightforwardly. In doing so
 an immaterial constant, associated with the first term in Eq.(\ref{te}) that
  contains the scattering length $a$, is absorbed into the chemical potential
   $\mu$ to yield
 \beq
   \epsilon(p)={p^2\over 2M^*}-\mu\  ,\quad {M\over M^*}=1-2\pi a^2r_e\rho  \  .
   \label{npt}
   \eeq
   Upon accounting for  a contribution $\sigma$ from long-wavelength spin fluctuations,
    relevant e.g. in the case of   3D liquid $^3$He discussed below,
   these formulas  change. Nevertheless, as it was shown  in microscopic calculations of
    the spectrum $\epsilon(p)$,  performed
   in Ref.\onlinecite{krotscheck},  the effective mass approximation remains adequate, merely
    the effective mass  is modified to
     \beq
     {M\over M^*}=1-2\pi a^2r_e\rho-\sigma  \  ,
     \label{meffs}
     \eeq
  with $\sigma\simeq 0.5$.    With these results, the temperature evolution of the
        single root $p_{1/2}(T)$  turns out to be identical to that in ideal Fermi gas.

   Temperature $T_M$, at which the chemical potential changes sign, is straightforwardly
  evaluated from the normalization condition (\ref{norm}) and formula  (\ref{npt}) to yield
   \beq
   T_M\simeq {p^2_F\over 2M^*(\rho)}   .
   \label{tm}
   \eeq
   As correlations are strengthening, the  ratio  $M^*/M$ increases  and consequently,
 temperature $T_M$   goes down.

 Unfortunately, in the vicinity of the QCP
  where the effective mass $M^*$  diverges,  the  effective mass  approximation
  begins  to fail, because other terms of the Taylor  expansion     come into  play.
    For illustration, let us  retain only    the next   term
 $\propto ({\bf p}_1-{\bf p}_2)^4$.  In  this case, the group velocity
 $d\epsilon/dp$ acquires the form
      \beq
 {d\epsilon(p)\over dp}=\alpha {p\over M}+\nu {p^3\over p^2_FM}
 \label{gr4}
 \eeq
 with $\nu>0$ and $\alpha$, being  dimensionless quantities, whose values are
 supposed to be small.  As long as $\alpha$  keeps positive sign,  the bifurcation
  momentum $p_b$ equals 0.    At the critical point where $\alpha$ vanishes,
 one has $ d\epsilon(p\to p_F)/dp\equiv p_F/M^*=\nu p_F/M$, so that $M/M^*=\nu\ll 1$.
 Repeating the same manipulations that give rise to Eq.(\ref{tm}), one then finds
  $T_M\propto \nu p^2_F/M$,    and   therefore $T_M/\epsilon^0_F\ll 1$.

  When $\alpha$ changes sign, the single-particle spectrum, found with the aid of
 simple integration of Eq.(\ref{gr4}),  can be conveniently rewritten in the form:
 \beq
 \epsilon(x)= \nu {p^2_F\over 4M}\left(x-x_b\right)^2-\mu(T) \  ,
 \label{e4}
   \eeq
 where $x=p^2/p^2_F$ and $x_b=|\alpha|/\nu$.  Again the effective mass  $M^*$ turns out
  to be enhanced: $M^*/M\simeq \nu^{-1}$.

The single root $p_F$ of $T=0$ equation  (\ref{topeq})
  persists as long as  $x_b<1/2$, otherwise, this equation acquires the second root
  $p_<(0)=p_F(2x_b-1)$, and the  Landau state is rearranged. The  single-particle
    states with $p_<<p<p_F$, where  $p_F$ is a new Fermi momentum, determined
  by the normalization condition (\ref{norm}),  remain filled, while states with $p<p_<$
   and $p>p_F$ turn out to be empty.    With the rise of temperature, both the roots of
Eq.(\ref{topeq}) move to meet each other at the bifurcation momentum $p_b=p_F\sqrt{x_b}$.
 Critical temperature $T_M$ is
 straightforwardly evaluated from Eq.(\ref{norm}). In the relevant case $x_b\simeq 1$,  one finds
 \beq
 T_M\simeq \nu p^2_F/M\ll \epsilon^0_F ,
 \label{tmm}
 \eeq
 implying again that the  elevation of temperature results in the rapid shrinkage of the FL domain.

 {\it Discussion.}
    All the analytically solved models, addressed above, have a common  feature: strengthening
 interactions results in the shrinkage of the temperature domain where standard FL results
 hold.   This  conclusion  remains valid
 even if  flattening
  of the single-particle   spectrum   $\epsilon(p)$ takes  place only for occupied states,
   the situation,  inherent in many strongly correlated systems where the
   spectrum $\epsilon(p)$ has a more complicated structure, ( see e.g. Ref.\onlinecite{prb2008}).
    The shrinkage becomes especially pronounced  beyond the QCP,\cite{zb,shagp,prb2008}
 where as mentioned above, the Fermi surface becomes  multi-connected. For illustration,  let Eq.(\ref{topeq}) have
 $n$ roots at $T=0$, implying that the Fermi surface has   $n$  sheets.
 Assuming the single-particle spectrum to be a
  parabolic function between two neighbour roots of Eq.(\ref{topeq}), one finds
  \beq
 \epsilon_{\rm max}\simeq {\epsilon_{QCP}\over n^2}  \  ,
 \eeq
   where $\epsilon_{\rm max}= {\rm max} |\epsilon(p)|$ inside the Fermi volume, i.e. for occupied states, and
   $\epsilon_{QCP}$ is the QCP Fermi energy. Since,  as seen from Eq.(\ref{dist}),  the critical collapse condition
    $n(p)<1/2$ is met  at   $T_M\simeq \epsilon_{\rm max}$, we infer that beyond the
QCP, critical  temperature   $T_M$  rapidly falls with increasing the number $n$ of sheets of the Fermi surface.

   Let us now briefly discuss the collapse of the Fermi surface in strongly correlated systems,
 where at $T=0$, there exist  flat bands. A different name for this dispersionless
 portion of  the single-particle spectrum $\epsilon(p)$ is the fermion condensate (FC).\cite{ks}
  As long as the FC density is small,  the collapse is insensitive to the presence of the FC.
  Although with further increasing the FC fraction, temperature $T_M$ rapidly falls, it  keeps a
  nonzero value, even if all the quasiparticles get to the FC. This conclusion holds  until
  $f_1$  attains a critical value, at which the inequality $n(p)<1/2$ is met for  any momentum $p$.
   In this situation, Eq.(\ref{topeq}) has no roots, and hence, such Fermi systems have {\it no Fermi surface
   even at $T=0$}.

   It is instructive to trace how the  standard FL regime that operates in
 the $T\to 0$ limit  gives way to a different one at $T>T_M$ where  Eq.(\ref{topeq})
 has no roots at all. For illustration, let us address the spin susceptibility
  $\chi(T)$, neglecting  for a while spin-spin correlations. In this case, $\chi(T)$
   is given by   the standard FL formula
 \beq
 \chi(T)\propto {1\over T}\int n(p)(1-n(p))d\upsilon  \  .
 \label{chit}
 \eeq
 In the FL regime, overwhelming contributions   to this integral come from a domain
 $|\epsilon(p)|\leq T$, adjacent to the Fermi   surface,  and after simple  algebra with
 employing the FL formula  $dp/d\epsilon=M^*/p_F$,  we arrive at the  Pauli result:
 $\chi(T)=const\propto M^*$.

   The collapse of the Fermi surface leads to the reconstruction of the FL regime.
 As $T$ increases away from $T_M$,  the chemical potential $\mu$ becomes more and more negative,
that  triggers exponential suppression   of the integral with the $n^2(p)$ term
  on the r.h.s. of Eq.(\ref{chit}). Eventually this contribution becomes negligible,
  and  the integral (\ref{chit}) is then evaluated in the closed form  with the aid
  of the  normalization condition (\ref{norm}), producing the classical   Curie law:
  $\chi(T)\propto \rho/T$.  Accounting for  the spin-spin correlations  introduces a
  correction, associated    with the  Weiss temperature $\Theta_W$, to yield
  $\chi(T)\propto \rho/(T-\Theta_W)$. Thus from the analysis of the spin susceptibility,
 we infer  that temperature $T_M$, at which the Fermi surface collapses,    does specify
 the  topological crossover from the standard FL regime to    a classical-like one.
 The verification of  this conclusion on the base of  experimental data on the specific
  heat $C(T)$ encounters difficulties,  requiring
 information on the spectrum $\epsilon(p,T)$ at $T>T_M$ over a wide momentum range that remains  scarce yet.

      3D liquid $^3$He  is one of  the systems that experiences  the topological
   crossover, discussed in this article. The   Fermi energy of this liquid
    is $\epsilon^0_F\simeq 5 K$, while the effective mass  $M^*(\rho)\geq 3M$.
    At very low temperatures,  behaviors of both $C(T)$ and $\chi(T)$    are known
  to  be consistent with FL theory.  However,    already at $T\simeq 0.2 K$, the spin
     susceptibility $\chi(T)$ begins to decline, and at $T\geq p^2_F/2M^*\simeq 1.5K$ its
      behavior is, indeed, obeys  the classical Curie   formula $\chi(T)\propto \rho/T$.
  Meanwhile  conventionally,  the departure of experimental data  from  the FL result
 $\chi(T)=const$  is   attributed to damping  effects.\cite{pines} Alas, how
  accounting for  damping  may lead to the Curie   behavior of $\chi(T)$ in liquid $^3$He,
   is hardly explicable.

 Let us now turn  to   strongly  correlated  electron systems   where the effective
 electron mass $M^*$ is  enhanced as well. To clear up confusion, caused by the presence
 of the lattice field in solids, it should be pointed out that it results in   a
  modification of Eq.(\ref{lansp}) that can be done on the base
   of   FL relation, ( see e.g. Ref.\onlinecite{migt}),
  \beq
  z{\cal T}^k({\bf p}) =z{\cal T}^{\omega}({\bf p})+
   \int f({\bf p},{\bf p}_1) {\partial n(p_1)\over \partial {\bf p}_1}{d^3p_1\over (2\pi)^3}   \  ,
  \eeq
  connecting  $\omega-$ and  $k$-limits of the vertex part ${\cal T}({\bf p})$,
  independently of whether the external field presents or not. In this equation,
 $z$ stands for    the renormalization factor.  The l.h.s. of this equation is
  expressed in the closed form:
  $z{\cal T}^k({\bf p})=-M\partial G^{-1}(\epsilon=0,{\bf p})/\partial {\bf p}=M\partial \epsilon(p)/\partial {\bf p}$.
   As for the quantity $z{\cal T}^{\omega}({\bf p})$, it can be written in the  Migdal form:
    \beq
    z{\cal T}^{\omega}({\bf p})=e_q{\bf p} \  ,
    \eeq
    where $e_q$ is the quasiparticle effective charge\cite{migt} that equals unity  in homogeneous
  matter due to momentum  conservation. In solids, momentum  conservation breaks down,
  and therefore $e_q\neq 1$. Additionally, in  solids, the electron mass $M$ is replaced
   by the   so-called LDA  electron mass $M_{LDA}$, which   cannot be evaluated with great
 precision because of the complicated    structure of the lattice field.  Therefore the
  effective charge   $e_q$ can be absorbed into $M_{LDA}$.  With this
  specification,    the above analysis can be applied to data on the magnetic susceptibility
 of the heavy-fermion  alloy Yb(Rh$_{1-x}$Co$_x)_2$Si$_2$, obtained at different dopings $x$
 in Ref.\onlinecite{steg2011}. Experimental results evidence for  {\it gradual}   flattening
 out of the product $T\chi(T)$, like in 3D liquid $^3$He.   However, the authors\cite{steg2011}  attribute such a behavior
to  the localization of  4$f$-electrons, notwithstanding  the localization is a phase   transition,
  rather than crossover, and  therefore the localization scenario fails to explain  gradual flattening of the function $T\chi(T)$.

 {\it Impact of the  collapse of the Fermi surface on pairing correlations}.
Phase transitions, associated with pairing correlations, are traditionally investigated
 on the base of  the BCS gap  equation. In the case of $s$-pairing, this equation reads
\beq
\Delta(p,T)=-\int {\cal V}( p,p_1) {\tanh {E(p_1)\over 2T}\over 2E(p_1)}\Delta(p_1,T)d\upsilon_1  \  .
\label{bcsd}
\eeq
 Here $\Delta(p)$ is the gap function,   $E(p)=\sqrt{\epsilon^2(p)+\Delta^2(p)}$ is
 the Bogoliubov  quasiparticle energy, and ${\cal V}( p, p_1)$ is the  zero harmonic
 of  the pairing   interaction  ${\cal V}({\bf p},{\bf p}_1)$, presented by a set of
 Feynman  diagrams, irreducible in the particle-particle channel.
  Upon neglecting its  momentum dependence  we are led  to the  well-known  BCS result:
   \beq
     \Delta(0)=\Omega_D e^{-2/\lambda} \  ,
     \label{d0}
    \eeq
   with the dimensionless pairing constant $\lambda=p_FM^*|{\cal V}_L(p_F,p_F)|/\pi^2$ and
   $\Omega_D$, the Debye frequency.

      Critical temperature $T^*$, at which pairing correlations die out, is found from
     homogeneous  equation
  \beq
D(p)=-\int {\cal V}( p,p_1) {\tanh {\epsilon(p_1,T^*)\over 2T^*}\over 2\epsilon(p_1,T^*)}D(p_1)d\upsilon_1  \ ,
\label{th}
\eeq
determining the location of the pole of the two-particle Green function at the
 total momentum ${\bf P}=0$, (Thouless criterion). Conventional wisdom reads that $T^*$ is
 critical temperature of termination of BCS superconductivity. Indeed, in the weak-coupling
 limit, Eq.(\ref{th})  is derived from Eq.(\ref{bcsd}), setting there $\Delta\to 0$,
  and therefore temperature   $T^*$  coincides with  BCS critical temperature $T_c$.  In
   this case, $T_c$ is expressed in terms of the gap magnitude  $\Delta(0)$ as
\beq
T_c=0.57 \Delta(0)  \ .
\label{tcb}
\eeq
However,  with  strengthening   correlations,   the   density of states, proportional to the
 effective mass $M^*$, increases, and both the quantities $\Delta(0)$ and  $T_c$ soar up, while
  $T_M$  goes down. Eventually, when inequality $T_M<T_c=0.57 \Delta(0)$ is met, the coincidence
between  $T^*$ and $T_c$  is destroyed. Indeed, at $T>T_M$, the  Gor'kov term
$\Delta^2/(\varepsilon+\epsilon(p))$ in the mass operator $\Sigma(p,\varepsilon)$  has no longer
 pole  at the Fermi surface, or equivalently at $\varepsilon\to 0$, due to  the fact that
  $\epsilon(p,T)>0$ at any momentum $p$. Therefore at $T>T_M$ the Gor'kov term has to be treated on equal
   footing with  regular contributions to $\Sigma$ that results in recovering the Dyson-like
 form of the quasiparticle Green function  $G(p,\varepsilon)=(\varepsilon-\epsilon(p))^{-1}$.
 In this situation, the Meissner  effect no longer exists, and  superconductivity
  is terminated.

  To clarify the structure of the pairing correlations  in the domain  $T_M\leq T<T^*$ where,
  according to Eq.(\ref{th}), these correlations  still persist, it is worth remembering the
  familiar situation with low-density symmetric nuclear matter at $T=0$ where the deuteron pole
   in the n-p scattering exhibits itself in full   force.
 In this case,  analyzed in numerous theoretical studies,\cite{eagles,leggett,nozs,roepke,lombardo}
 the  pairing correction to the chemical potential $\mu(0)$ turns  out to be in great  excess
 of   the Fermi energy $ p^2_F/2M$. As a result, the  Cooper  condensate  transforms
into a Bose condensate of   bound states of  quasiparticle pairs, the boson  radius $r_B$ being
 much less than the interparticle distance $r_0$, (the BCS-BEC crossover).   Analogous results have been
  obtained  in the model of bipolaronic superconductivity,\cite{alex1} where the strength of the
electron-phonon attraction is supposed to be large: $\lambda\gg 1$.   With decreasing  density $\rho$ or increasing
 $\lambda$, the   chemical
 potential $\mu(0)$ diminishes and eventually becomes negative,   as   in  the classical situation: $\mu(0)\to -\epsilon_o/2$
 where  $\epsilon_o$  stands for the quasimolecule  binding energy.

 Above critical  temperature $T_B$, at which the Bose condensate dies out, a conglomerate of  moving
  bound pairs remains that coexists with ordinary quasiparticles, the well-known feature of hot
  low-density nuclear matter. This coexistence   manifests itself in the  enhancement of deuteron
 formation  in deep inelastic nuclear reactions. In unison  with this fact, a large diamagnetic
 response,  observed in  high-$T_c$ superconductors, is explained, attributing it to the
  normal-state Landau diamagnetism of the bipolaron system.\cite{alexd}

 In contrast to the case of low-density matter, in dealing with strongly correlated Fermi systems, the energy $\epsilon_o$,
   determined by the Bethe-Salpeter equation
  \beq
D(p)=-\int {\cal V}( p,p_1){\tanh {\epsilon(p_1,T)\over 2T}\over 2\epsilon(p_1,T)-\epsilon_o(T)} D(p_1)d\upsilon_1  ,
\label{ord}
\eeq
depends on   temperature, and  as stems from comparison of  Eqs.(\ref{th})  and (\ref{ord}),
   $\epsilon_o(T)$ vanishes at $T=T^*$.

 Our first goal is to evaluate a critical point of  the Lifshitz phase diagram for systems with pairing correlations,
  where both the Cooper condensate and the conglomerate of bound
 pairs disappear simultaneously that occurs provided temperatures $T_M$ and $T^*$ coincide with each other.
 The corresponding critical value $\lambda_c$ is  found with the  aid of the Thouless criterion
 (\ref{th}). In doing so we employ the above model  with the single-particle  spectrum (\ref{e4})
 where the root   $p_{1/2}(T_M)$ is assumed  to be located quite far from  the origin:
 $p_{1/2}(T_M)/p_F\simeq 1$.  Remembering   that at $T=T_M\propto 1/M^*$, the chemical
  potential $\mu(T)$   vanishes,  we arrive at         the following  equation
  \beq
  1\propto {\lambda_c\sqrt{{\epsilon^0_F\over M^* T^*}}}\int\limits_0^{\infty}  {\tanh z\over z^{3/2}}dz   \ ,
\label{tcm}
  \eeq
 to find that in the case $T^*=T_M$,
  \beq
\lambda_c\simeq O(1)  .
 \label{crr}
 \eeq

  This  result holds    beyond the QCP as well. Indeed, by definition,    all  new pockets
   of the Fermi surface that emerge beyond the QCP disappear at $T=T_M$. Near the
  bifurcation momentum $p_{1/2}(T_M)$, the  single-particle spectrum $\epsilon(p,T_M)$ has
  the  same   parabolic  form:   $\epsilon(p,T_M)\propto w(p-p_{1/2}(T_M))^2$ as in
 Eq.(\ref{e4}), with the prefactor $w\propto 1/M^*$. Evidently, upon inserting these relations
  into  Eq.(\ref{th}) the    result   (\ref{crr}) is recovered.

\begin{figure}[t]
\includegraphics[width=0.95\linewidth,height=.68\linewidth]{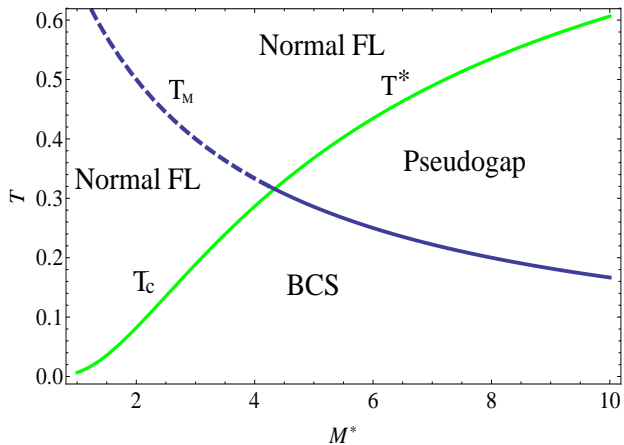}
\caption{Color online.
The phase diagram of a Fermi system with pairing correlations vs. the effective mass $M^*$.
The strength of   pairing interaction ${\cal V}$ is supposed to be small
 enough to avoid the regime of bipolaronic superconductivity.
 The line, separated the state with pairing correlations  from the conventional  FL state, is drawn in green.
 The pseudogap phase is separated from the BCS one by  the solid blue line  where the collapse
 of the Fermi surface occurs.
 The dashed blue line separates the Landau states from the normal state without the Fermi surface.
 Both temperature $T$
 and the effective mass $M^*$ are given in arbitrary units.
 }
\label{fig:bcp}
\end{figure}

  Evidently, in case  $M^*$ increases,  temperature $T_M$ drops, while
   $T^*$ grows. Thus in the domain of the phase diagram where $\lambda\simeq 1$,
   the BCS superconducting phase transition is split into two ones.  The lower transition continues to  be associated with
   termination of superconductivity, while  the upper one turns out to be  related to  the extinguishment of  the bound pairs.

  The excess $T^*-T_M$ can be  shown to obey  a  linear relation
 \beq
T^*-T_M=T_M(\lambda-\lambda_c) \  ,
\eeq
 determining the range of the region $T_M<T<T^*$ where the quasiparticles  coexist with
 the quasimolecules, while the BCS correlations are completely suppressed.

 It is instructive to evaluate the binding energy $\epsilon_o(\lambda,T)$ at  critical
 temperature $T_M$.  At small difference $(T^*-T_M)<<T^*$, this energy is
 calculated from a set of two equations, one of which is Eq.(\ref{tcm}), while the
 second reads
  \beq
  1\propto {\lambda\sqrt{{\epsilon^0_F\over M^* T_M}}}\int\limits_0^{\infty}  {\tanh z\over z^{1/2}(z+\epsilon_o/T_M)}dz   \ .
\label{tr}
  \eeq
 Straightforward calculations then yield
 \beq
 \epsilon_o(\lambda,T_M)\propto T_M (\lambda-\lambda_c)^2 \  .
 \eeq
 One of consequences of this result is that the pairing correlations cannot regain its
 original BCS form at $T=T_M$, and therefore the BCS critical temperature $T_c$ must be lower
  than $T_M$.  The analysis of the interplay between the two types of these correlations
  at low $T<T_M$, dating back to Ref.\onlinecite{eagles}, is not properly   performed yet.
  I hope to  revert to  this question in the future.
 These results are summarized in   the Lifshitz  phase diagram of a  Fermi
  system with pairing correlations, drawn in  Fig.~\ref{fig:bcp} where the effective mass $M^*$ changes, while
   the pairing interaction ${\cal V}$ remains  unchanged.

   Let us now  turn to pairing correlations in  systems with  flat bands, being nowadays the
    subject of   wide speculation because of  the   huge density of states of these
  systems.\cite{vol2011b,vol2011c,kopnin}   As we have seen, if in a system with the flat band,
  there exists  the point $p_{1/2}$  where $n(p)=1/2$, such a system possesses the Fermi surface,
  and one  can expect that its  superconductivity is  described within BCS theory.
  In systems without  the  Fermi surface where  inequality $n(p)<1/2$ is met  at any point of
  momentum space,   the $T=0$ ground state of the system with the flat band, being superfluid,
  is a condensate of the bound  quasiparticle pairs.

  In conclusion, in this article, topological crossover in homogeneous matter, associaited with
   the collapse of the Fermi surface, is analyzed.  It is shown that this collapse leads to
    the replacement of   the standard FL regime by a classical Maxwell-like one.  Temperature
    behavior of the spin susceptibility of 3D liquid $^3$He is demonstrated  to be properly
    explained  within this  scenario     that implies  the substantial extention of the
 temperature region where the quasiparticle pattern of phenomena  in 3D liquid $^3$He holds.
    The impact of the collapse of the Fermi surface on pairing   correlations in a  domain
 of the Lifshitz phase diagram
  is examined. The topological crossover, connected  with the collapse, is shown  to be
  responsible   for  splitting of the BCS superconducting phase transition into two ones of
 {\it the same symmetry}. The lower one is related to termination of BCS superconductivity,
 while the upper one is  associated with the disapperance  of the conglomerate of the  bound
 pairs, a key feature of  the pseudogap phenomenon.\cite{shen}

       I thank A. Alexandrov, A. Bratkovski, J. Clark, H. Godfrin, P. Schuck, V. Shaginyan    and
  M. Zverev for fruitful    discussions.  This research was supported by  the McDonnell Center
   for the Space Sciences, by  Grants No.~2.1.1/4540 and NS-7235.2010.2 from the Russian
Ministry of Education and Science, and by Grant No.~09-02-01284
from the Russian Foundation for Basic Research.
   
\end{document}